\documentclass[twocolumn,showpacs]{revtex4}
\usepackage{dcolumn}
\usepackage{amsmath}
\makeatletter
\providecommand{\LyX}{L\kern-.1667em\lower.25em\hbox{Y}\kern-.125emX\@}

\usepackage[T1]{fontenc}
\usepackage[latin1]{inputenc}
\usepackage{graphics}
\makeatother

\begin{document}

\title{Tensor representation of the nucleon-nucleon amplitude}

\author{R. Crespo and A.M. Moro}.

\affiliation{
Departamento de F\'{\i}sica, Instituto Superior T\'ecnico, 
Av.\ Rovisco Pais, 1049-001, Lisboa, Portugal 
}

\date{\today}

\newcommand{\be}{\begin{eqnarray}}
\newcommand{\ee}{\end{eqnarray}}

\begin{abstract}
Many approaches to nucleon-nucleus elastic and inelastic scattering are
based on the use of the free-space nucleon-nucleon transition
amplitude. In calculations where the full spin-dependence of this
amplitude is needed, its use is more tractable when it is expressed in
terms of irreducible tensor operators of the spins of the interacting
nucleons. We present general  formulae for this representation
which is particularly useful for inelastic scattering studies
involving spin-flip transitions of a target nucleon.
\end{abstract}

\pacs{PACS catagories: 24.10.--i, 24.10.Ht, 24.70.+s, 25.40.Cm}

\maketitle

\section{Introduction}

The free nucleon-nucleon (NN) transition amplitude, both on- and off-
the energy shell \cite{A,B} is a basic dynamical input off
{\em ab initio} multiple scattering
scattering formalisms either within a many-body \cite{KMT,Watson}
or few-body \cite{MST} framework.
These formalisms have been applied with success to describe elastic
and inelastic scattering from stable \cite{CJT}(and references therein)
and unstable nuclei \cite{MSTclus,evora}. 
The  spin  and isospin dependent amplitudes of the transition amplitude
can be calculated readily from realistic NN interaction models,
such as the Paris \cite{Paris} and Bonn \cite{Bonn} interactions.
It was shown \cite{B} that the off-shell behaviour of the NN
transition amplitude is very stable against the underlying NN interaction.
In addition,  the on- shell values must reproduce the available 
experimental NN data. Thus, the dynamical NN input of the multiple
scattering expansions is very well defined at least on- the energy shell.

Traditionally, the NN transition amplitude has been presented using the
Wolfenstein parametrization \cite{WOL}. This representation involves
six amplitudes ${\cal A},\ldots {\cal F}$, being the coefficients of
spin operators which are scalar products of the Pauli spin vectors
$\vec{\sigma}_i$ for the projectile and struck nucleons with a set of
unit vectors defined by the scattering plane of the nucleon pair.  This
representation is not convenient in cases, such as arise for example
in inelastic scattering, where one needs to account fully for the the
spin dependence of the NN interaction \cite{Ed}.
A more tractable representation for treating this spin dependence 
is presented here.

\section{Formalism}
Assuming that we use plane wave states, normalized such that
\be
\langle \vec{r}\,|\vec{k}\rangle  = (2\pi)^{-\frac{3}{2}} \exp(i\vec{k}
\cdot\vec{r}\,)~,
\ee
then the free NN scattering amplitude $M(\omega,\vec{\cal K}',
\vec{\cal K})$, describing scattering from two-nucleon states with
relative momenta $\vec{\cal K}$ and $\vec{\cal K}'$ for relative energy
$\omega$ in their centre of mass (cm) frame, is related to the
anti-symmetrised transition matrix elements by
\be
M(\omega,\vec{\cal K}',\vec{\cal K})=\langle \vec{\cal K}'| M(\omega)|
\vec{\cal K}\rangle  = -\frac{4\pi^2\mu}{\hbar^2} \langle \vec{\cal K}'|
t^f_{01}(\omega)|\vec{\cal K}\rangle ~, \label{scattA}
\ee
where $\mu$ the NN reduced mass. These amplitudes are operators in
both the NN spin  and isospin spaces.

The Wolfenstein decomposition of the NN amplitude for the scattering of
an incident (0) and struck (1) nucleon has been used extensively.  It
writes the most general form of the amplitude, consistent with
time-reversal, parity, and  rotational invariance, as
\be
M(\omega,\vec{\cal K}',\vec{\cal K})&=&{\cal A} + {\cal B}
(\vec\sigma_0\cdot\hat{n})(\vec\sigma_1\cdot\hat{n})
+ {\cal C}(\vec\sigma_0+\vec\sigma_1)\cdot\hat{n}
\nonumber \\ &+& {\cal D}
(\vec\sigma_0\cdot\hat{m}) (\vec\sigma_1\cdot\hat{m}) +
{\cal E} (\vec\sigma_0\cdot\hat{\ell}\,)
(\vec\sigma_1\cdot\hat{\ell}\,)
\nonumber \\ &+&{\cal F}[
(\vec\sigma_0\cdot\hat{\ell}\,)(\vec\sigma_1\cdot\hat{m}) +
(\vec\sigma_1\cdot\hat{m})(\vec\sigma_0\cdot\hat{\ell}\,)] \label{KMT}
\ee
where the orthogonal set of unit vectors $\hat{n}=(\vec{\cal K}\times
\vec{\cal K}')/|\vec{\cal K}\times\vec{\cal K}'|$,
$\hat{\ell}=(\vec{\cal K}' +\vec{\cal K})/|\vec{\cal K}'+\vec{\cal
K}|$, and $\hat{m}=\hat{\ell} \times\hat{n}$ are defined by the NN
scattering plane \cite{WOL}. The coefficient amplitudes ${\cal A},
\ldots {\cal F}$ can also be expressed as complex functions of
$\omega$, the momentum transfer $\vec{q}=(\vec{ \cal K}'-\vec{\cal K})$
and the total momentum $\vec{\cal Q}=(\vec{\cal K}'+ \vec{\cal K})/2$
of the NN pair in their cm frame. They remain operators in isotopic
spin space, so for instance
\be
{\cal A}(\omega,\vec{q},\vec{\cal Q})&=&{\cal A}_0 + {\cal A}_\tau
(\vec{\tau}_0 \cdot \vec{\tau}_1)\nonumber \\ &=&{\cal A}^{T=0}P_0 +
{\cal A}^{T=1}P_1 ~~,\label{tautau}
\ee
where the $\vec{\tau}_i$ are the isospin Pauli operators for the two
nucleons and the $P_T$ are projectors for the stated total isospin
states. In this equation, ${\cal A}_0$ and ${\cal A}_\tau$ represent the
isoscalar and isovector componenents of ${\cal A}$ in the isospin space.

The ${\cal A} \cdots {\cal F}$ amplitudes can be derived on- and off- the
energy shell \cite{A,B,CJT} from a realistic NN interaction such as 
for example the Paris \cite{Paris} and Bonn \cite{Bonn} interactions. 

Equivalently, the NN scattering amplitude can be represented in terms
of central, spin-orbit and tensor components. 
The work of Franey--Love (FL) \cite{Love} makes use of this
representation, where the form factors of the components
are non local, with a direct and an exchange term
and  given in terms of sums of Yukawa forms,
the ranges and strenght of each term being determined by
fitting the on-shell NN scattering data. 
The FL transition  amplitude is usually referred as a pseudo-T-matrix
since it is not constructed from a potential model
and violates unitarity badly \cite{Mac88}.

In the example of nucleon elastic scattering on a spin-zero target 
only the central, ${\cal A}$, and spin-orbit, ${\cal C}$,
components contribute to the
first order term of the multiple scattering expansion of the optical
potential \cite{KMT,Watson}. Second and higher order terms, however,
involve contributions from all components \cite{CJT}.
Of course, for elastic scattering from a non spin-zero nucleus,
or for inelastic scattering involving spin-flip transitions
of a struck nucleon in the target, a full treatment of the scattering 
amplitude needs to be considered.
In these applications, approximate treatments need to be performed
to  handle the orthogonal set of unit vectors, 
 ${n}=(\vec{k}\times \vec{k}')/|\vec{k}\times\vec{k}'|$, 
${\ell}=(\vec{k}' +\vec{k})/|\vec{k}'+\vec{k}|$,
and ${m}={\ell} \times{n}$ \cite{KAW,CJT}.

In elastic scattering processes involving 
a more general target, and in inelastic scattering, in which the
full spin dependence of the interacting nucleons enters, the NN
amplitudes are more conveniently constructed such that
\be
\langle \vec{\cal K}'| M | \vec{\cal K}\rangle  = \sum_{a\alpha ~ b\beta}
M^{(ab)}_{\alpha \beta }(\vec{\cal K}',\vec{\cal K})
\tau_{a\alpha}(s_0) \tau_{b\beta}(s_1)
~~,
\ee
where $\tau_{a\alpha}(s_0)$ is the irreducible tensor operator for
the projectile particle (0) with spin $s_0$ ($a=0,\ldots 2s_0$); 
$\tau_{b\beta}(s_1)$ is the irreducible tensor operator for
the struck particle (1) with spin $s_1$ ($b=0,\ldots 2s_1$).
Explicitly, since $s_0 = s_1={\scriptstyle\frac{1}{2}}$,
\be
\tau_{00}({\scriptstyle\frac{1}{2}}) = 1~,~~~~~~~\tau_{1\beta}
({\scriptstyle\frac{1}{2}}) =\sigma_\beta(1)~~,
\ee
with $\sigma_\beta(1)$ the spherical components of $\vec{\sigma}_1$
with respect to the chosen $z$-axis. It is understood that the
$M^{(ab)}_{\alpha \beta }$ depend on the isospin of the two nucleons. 
The amplitudes relevant to the pp, pn, and nn cases are obtained from the 
isospin singlet ($T=0$) and triplet ($T=1$) amplitudes.  The explicit
dependence of the $M$ and $M^{(ab)}_{\alpha \beta }$ on $\omega$ and $T$ will
not be shown in the following.

\begin{figure}
{\par\centering \resizebox*{0.45\textwidth}{!}
{\includegraphics{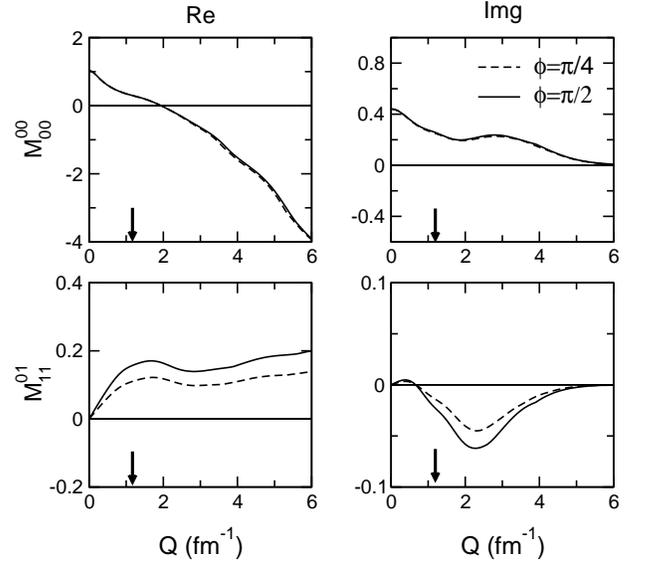}} \par}
\caption{\label{Fig:phidep} Real and Imaginary part of the
Isoescalar components of $M_{00}^{00}$ and $M^{01}_{11}$,
at E$_{\rm Lab}$= 135 MeV, as a function of the total momentum ${\cal Q}$, 
with q = 1 fm$^{-1}$. 
 }
\end{figure}

\begin{figure}
{\par\centering \resizebox*{0.45\textwidth}{!}
{\includegraphics{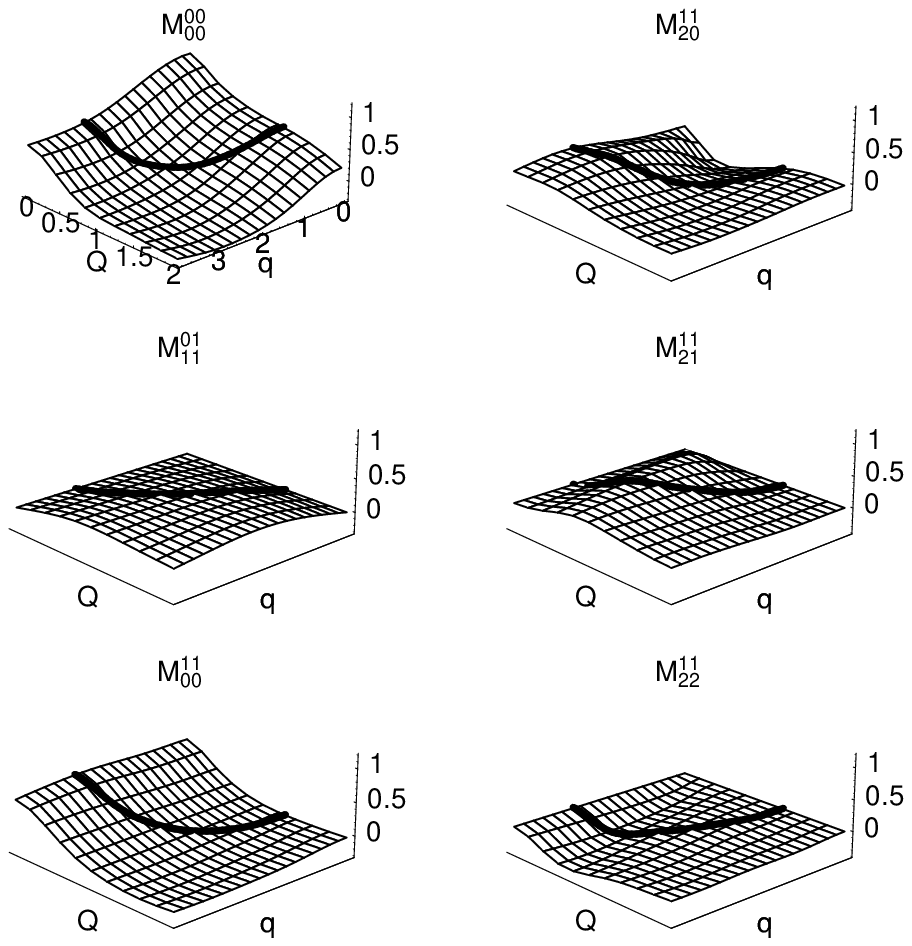}} \par}
\caption{\label{Fig:isoescalarRe} Real part of the 
Isoescalar components at E$_{\rm Lab}$= 135 MeV
 }
\end{figure}

We first decompose the NN amplitude of Eq.\ (\ref{scattA}) into spin
singlet ($S=0$) and triplet ($S=1$) components, $M^S_{\nu' \nu}$, where
$\nu$ and $\nu'$ refer to the incident and final state spin projections
in state $S$
\be
\langle \vec{\cal K}'| M | \vec{\cal K}\rangle = \sum_{S \nu
\nu'}M^{S}_{\nu'\nu}(\vec{\cal K}', \vec{\cal K}) |S\nu'\rangle
\langle S\nu|~.
\ee
These $M^S_{\nu'\nu}=\langle\vec{\cal K}'S \nu'| M | \vec{\cal K}S \nu
\rangle$ are in any case calculated during the construction of the NN
amplitudes from the partial wave transition amplitudes $M^{JS}_{L'L}(
{\cal K}',{\cal K})$, e.g. Appendix C of \cite{CJT}. We adopt the
convention that
\be
\langle \vec{\cal K}'| M | \vec{\cal K}\rangle &=& \frac{2}{\pi} \sum_{
JLL'SM}i^{L-L'} {\cal Y}^M_{(L'S)J}(\hat{\cal K}') M^{JS}_{L'L}(
{\cal K}',{\cal K}) \nonumber \\ 
&\times&{\cal Y}^{M~\dagger}_{(LS)J}(\hat{\cal K})~,
\ee
where ${\cal Y}^M_{(LS)J}$ is a spin-angle function
\be
{\cal Y}^M_{(LS)J}(\hat{\cal K}') = \sum_{\Lambda\nu}(L\Lambda
S\nu|JM)Y_{L\Lambda}(\hat{\cal K}'){\cal X}_{S\nu}~,
\ee
and $Y_{L\Lambda}$ and ${\cal X}_{S\nu}$ are spherical harmonics
\cite{Brink} and total spinors of the NN pair. Explicitly therefore
\be
M^S_{\nu'\nu} (\vec{\cal K}',\vec{\cal K})
 &=& \frac{2}{\pi}\sum_{JMLL'\Lambda\Lambda'}i^{L-L'}(L'\Lambda'
S\nu'|JM) \nonumber \\   & \times&  (L\Lambda S\nu|JM)
Y_{L'\Lambda'}(\hat{\cal
K}')Y^*_{L\Lambda}(\hat{\cal K})
\nonumber \\
&\times& M^{JS}_{L'L}({\cal K}',{\cal K}).
\label{msnn}
\ee
The partial wave sums are, of course, over values which satisfy the
Pauli principle requirement, $L+S+T$=odd. 

It is now convenient to reexpress the spin-space projector in terms
of irreducible tensor operators in the space of spin $S$ \cite{Ron},
\be
\tau_{\kappa q}(S) = \sum_{\nu\nu'} \hat{\kappa}
(S\nu\kappa q |S \nu') |S\nu'\rangle \langle S\nu|~,
\ee
where $0\leq \kappa \leq 2S$, and $\hat{\kappa} = \sqrt{2{\kappa}+1}$
which satisfy $\tau^\dagger_{\kappa q}(S) = (-)^q \tau_{\kappa -q}(S)$,
and so the spin tensor decomposition of the amplitude is
\be
\langle \vec{\cal K}'| M | \vec{\cal K}\rangle  = \sum_{Skq}
M^{S}_{\kappa q}(\vec{\cal K}',\vec{\cal K}) \tau^\dagger_{\kappa q}(S)
\ee
where
\be
M^{S}_{\kappa q}(\vec{\cal K}',\vec{\cal K}) =\frac{\hat{\kappa}}
{\hat{S}^2}\sum_{\nu\nu'} M^S_{\nu'\nu} (\vec{\cal K}',\vec{\cal K})
(S\nu' \kappa q |S \nu)~.
\ee

\begin{figure}
{\par\centering \resizebox*{0.45\textwidth}{!}
{\includegraphics{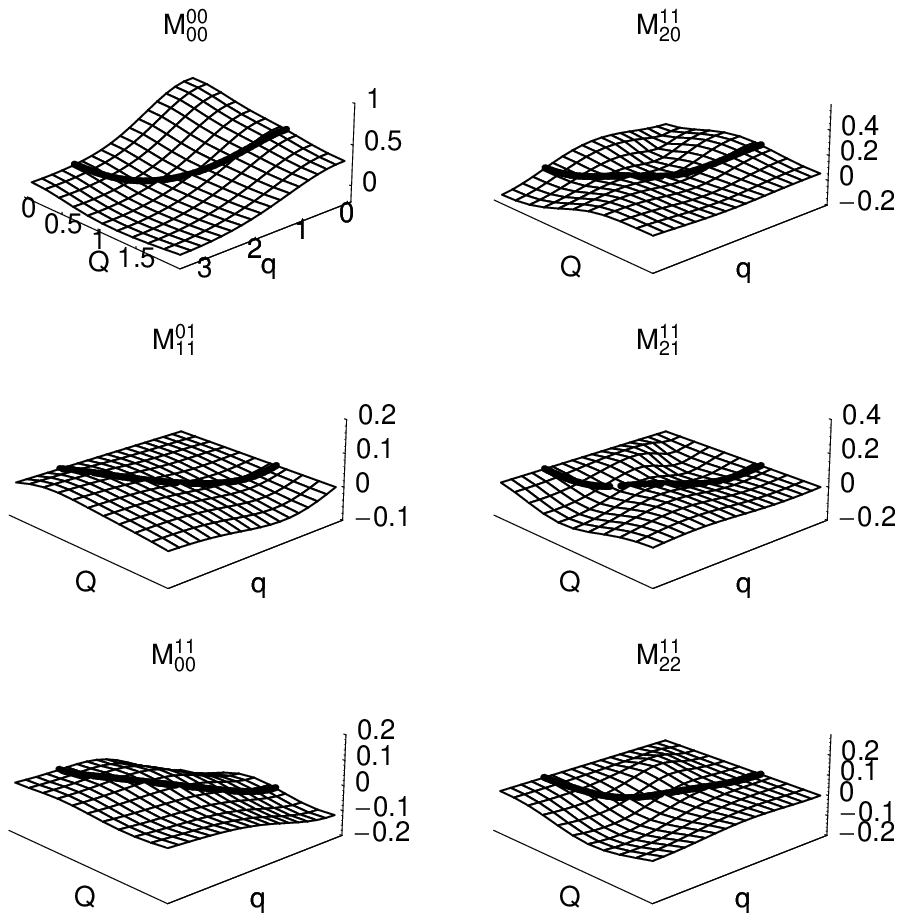}} \par}
\caption{\label{Fig:isoescalarIm} Imaginary part of the
Isoescalar components at E$_{\rm Lab}$= 135 MeV
 }
\end{figure}
It just remains to decompose the spin tensors for spin $S$ in terms
of those, $\tau_{b\beta}(s_i)\ (i=0,1)$, of the incident (0) and struck
(1) nucleons, as
\be
\tau_{\kappa q}(S) &= &\sum_{ab} \frac{ \hat{S}^3 \hat{a} \hat{b}}
{\hat{s}_0 \hat{s}_1} \left\{
\begin{array}{c c c}
s_0  & s_1 &  S \\
s_0  & s_1 &  S \\
a    & b   & \kappa
\end{array}
\right\} \nonumber \\ 
&\times &  \sum_{\alpha \beta} 
 (a\alpha b\beta |\kappa q) \tau_{a\alpha}(s_0) \tau_{b\beta}
(s_1)~~.
\ee
This, can be rewritten as
\be
\tau_{\kappa q}(S)  &=& \sum_{ab} \frac{ \hat{S}^3 \hat{a} \hat{b}}
{\hat{s}_0 \hat{s}_1} \left\{
\begin{array}{c c c}
s_0  & s_1 &  S \\
s_0  & s_1 &  S \\
a    & b   & \kappa
\end{array}
\right\}
{\cal T}_{\kappa q}(a,b)~,
\ee
where we have defined the new tensor operators
\be
{\cal T}_{\kappa q}(a,b) = \sum_{\alpha\beta} (a\alpha b\beta|
\kappa q) \tau_{a\alpha}(s_0) \tau_{b\beta}(s_1)~~.\label{bigT}
\ee
In terms of these,
\be
\langle \vec{\cal K}'| M | \vec{\cal K}\rangle  = \sum_{\kappa q~ab}
M^{(ab)}_{\kappa q}(\vec{\cal K}',\vec{\cal K}) {\cal T}^\dagger_{
\kappa q}(a,b)~~,
\ee
where $(a,b)$ refer to the ranks of the nucleon spin tensors, and
\be
M^{(ab)}_{\kappa q}(\vec{\cal K}',\vec{\cal K}) = \sum_{S} M^{S}_{
\kappa q}(\vec{\cal K}',\vec{\cal K}) {\cal N}^{S}_{\kappa }(ab)
\ee
with
\begin{equation}
{\cal N}^{S}_{\kappa }(ab) = \frac{ \hat{S}^3 \hat{a} \hat{b}}
{\hat{s}_0 \hat{s}_1}\left\{\begin{array}{ccc}
s_0  & s_1 &  S \\
s_0  & s_1 &  S \\
a    & b   & \kappa
\end{array} \right\} ~~.
\end{equation}
The amplitudes $M^{(ab)}_{\kappa q}$ can be expressed in terms of the 
transfered momentum  $q$, total momentum ${\cal Q}$, and the angle $\phi$
between these two vectors, that is, 
$ M^{(ab)}_{\kappa q}(\omega, q,{\cal Q},\phi)$. On- the energy shell,
$ \phi = \pi/2$ and $q^2/4 + {\cal Q}^2 = 2 \mu \omega / \hbar^2 $.

For checking purposes, we note that $M^{(00)}_{00}={\cal A}$, and
$M^{(01)}_{11}=M^{(10)}_{11}=-i{\cal C}/{\sqrt{2}}$.

We need to consider the non-vanishing amplitudes $M^{(00)}_{00}$,
$M^{(01)}_{11}=M^{(10)}_{11}$, $M^{(11)}_{00}$, $M^{(11)}_{20}$,
$M^{(11)}_{21}$, and $M^{(11)}_{22}$. The amplitudes also satisfy
$M^{(ab)}_{\kappa-q}= (-1)^{\kappa+q}M^{(ab)}_{\kappa q}$. All
terms with $a=b=\kappa=1$ are seen to be zero as a result the
vanishing of the 9-j coefficient.
As a result of this geometric coefficient, the tensor ${\cal T}_{\kappa q}$
satisfies ${\cal T}^+_{\kappa q} = (-1)^q{\cal T}_{\kappa -q}$.

\begin{figure}
{\par\centering \resizebox*{0.45\textwidth}{!}
{\includegraphics{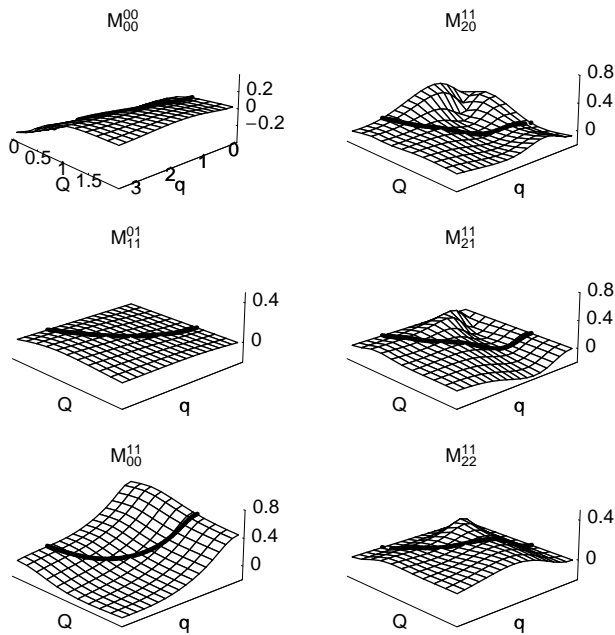}} \par}
\caption{\label{Fig:isovectorRe} Real part of the
Isovector components at E$_{\rm Lab}$= 135 MeV
 }
\end{figure}

\begin{figure}
{\par\centering \resizebox*{0.45\textwidth}{!}
{\includegraphics{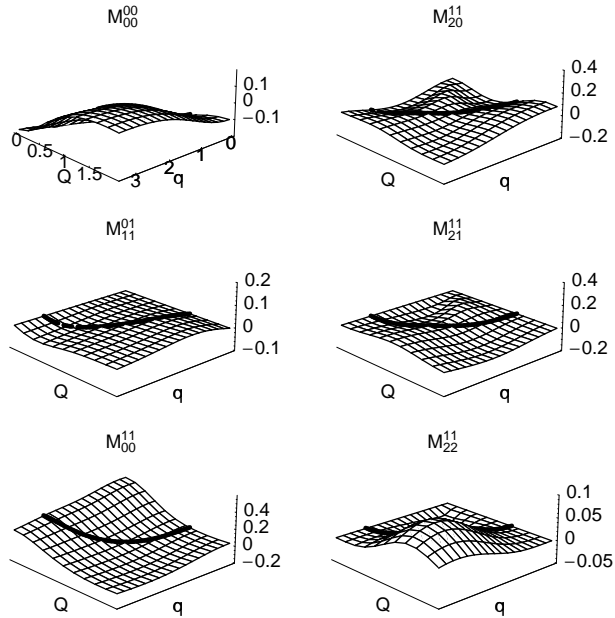}} \par}
\caption{\label{Fig:isovectorIm} Imaginary part of the
Isovector components at E$_{\rm Lab}$= 135 MeV
 }
\end{figure}

\section{Results}

For illustrative purposes, we show in here the calculated 
off- shell amplitudes, $M^{(ab)}_{\kappa q}$, at E$_{\rm lab}$=135 MeV,
making use of the Paris potential \cite{Paris}.
We  use NNAMP \cite{NNAMP} which calculates all the Wolfenstein and tensor
representation amplitudes on- and off- the energy shell.
The tensor amplitudes,  $M^{(ab)}_{\kappa q}$, are evaluated from 
the angular momentum amplitudes,
$ M^{JS}_{L'L}$, following the procedure described in the text.
These angular momentum amplitudes are obtained as in the work of
\cite{A,NNredish}. For this particular example, we have used a maximum
number of six partial waves. 

We have verified that the tensor amplitudes
do not vary strongly with energy and with the angle $\phi$.
From all the amplitudes, only  $M_{11}^{00}$ 
(and thus the Wolfenstein amplitude ${\cal C}$)
represented  in Fig.~\ref{Fig:phidep},
shows a slight dependence on $\phi$, the other amplitudes
remaining fairly independent on this parameter. In this figure,
the calculated amplitudes
are evaluated at q = 1 fm$^{-1}$, the solid and the
dashed lines corresponding to $\phi = \pi/2$ and $\phi = \pi/4$ respectively.
The arrow indicates the on- shell value.

In Figs.~\ref{Fig:isoescalarRe} and ~\ref{Fig:isoescalarIm}, 
we represent the real and imaginary parts of the isoescalar components
in the isospin space.
In Fig.~\ref{Fig:isovectorRe} and Fig.~\ref{Fig:isovectorIm} we show
the corresponding isovector components.

The amplitudes are represented as a function
of the  transfered momentum $q$ and the
total momentum ${\cal Q}$. The angle $\phi$
between these two vectors taken to its on- shell value, $\phi = \pi/2$.
The axis of quantization is chosen in the direction of the
incident beam.
The curve represented in each 3D plot represents the corresponding
on- shell value. 

It follows from the figures that effect of the nonlocalities 
for the rank 0  components of the tensor representation, 
$M^{(00)}_{00}$ and  $M_{(00)}^{11}$, might be significant when used in
multiple  scattering frameworks.
The amplitudes in other reference systems, can be
readily obtained from these, through rotation.

\section{Conclusion}

In summary, we have described  a convenient general method
to express the NN transition amplitude as a linear
combination for the spherical components of the spin operators of
the two interacting particles.
This is a more treatable representation to be used in multiple scattering
formalisms which required a full treatment of the spin of the NN transition 
amplitude.

\bigskip 
\bigskip

{\bf Acknowledgements:}

The authors would like to thank J.A. Tostevin and R.C. Johnson
for many usefull discussions.
This work was supported by Funda\c c\~ao para a Ci\^encia e Tecnologia
(Portugal) through grant No. POCTI/1999/FIS/36282.

\end{document}